# Variable Very High Energy Gamma-ray Emission from the Microquasar LS I +61 303

J. Albert[a], E. Aliu[b], H. Anderhub[c], P. Antoranz[d], A. Armada[b], M. Asensio[d], C. Baixeras[e], J. A. Barrio[d], M. Bartelt[f], H. Bartko[g], D. Bastieri[h], S. R. Bavikadi[i], W. Bednarek[j], K. Berger[a], C. Bigongiari[h], A. Biland[c], E. Bisesi[i], R. K. Bock[g], P. Bordas[u], V. Bosch-Ramon[u], T. Bretz[a], I. Britvitch[c], M. Camara[d], E. Carmona[g], A. Chilingarian[k], S. Ciprini[l], J. A. Coarasa[g], S. Commichau[c], J. L. Contreras[d], J. Cortina[b], V. Curtef[f], V. Danielyan[k], F. Dazzi[h], A. De Angelis[i], R. de los Reyes[d], B. De Lotto[i], E. Domingo-Santamaría[b], D. Dorner[a], M. Doro[h], M. Errando[b], M. Fagiolini[o], D. Ferenc[n], E. Fernández[b], R. Firpo[b], J. Flix[b], M. V. Fonseca[d], L. Font[e], M. Fuchs[g], N. Galante[o], M. Garczarczyk[g], M. Gaug[h], M. Giller[j], F. Goebel[g], D. Hakobyan[k], M. Hayashida[g], T. Hengstebeck[m], D. Höhne[a], J. Hose[g], C. C. Hsu[g], P. G. Isar[g], P. Jacon[j], O. Kalekin[m], R. Kosyra[g], D. Kranich[c,n], M. Laatiaoui[g], A. Laille[n], T. Lenisa[i], P. Liebing[g], E. Lindfors[l], S. Lombardi[h], F. Longo[p], J. López[b], M. López[d], E. Lorenz[c,g], F. Lucarelli[d], P. Majumdar[g], G. Maneva[q], K. Mannheim[a], O. Mansutti[i], M. Mariotti[h], M. Martínez[b], K. Mase[g], D. Mazin[g], C. Merck[g], M. Meucci[o], M. Meyer[a], J. M. Miranda[d], R. Mirzoyan[g], S. Mizobuchi[g], A. Moralejo[b], K. Nilsson[l], E. Oña-Wilhelmi[b], R. Orduña[e], N. Otte[g], I. Oya[d], D. Paneque[g], R. Paoletti[o], J. M. Paredes[u], M. Pasanen[l], D. Pascoli[h], F. Pauss[c], N. Pavel[m,x], R. Pegna[o], M. Persic[r], L. Peruzzo[h], A. Piccioli[o], M. Poller[a], G. Pooley[w], E. Prandini[h], A. Raymers[k], W. Rhode[f], M. Ribó[u], J. Rico[b†], B. Riegel[a], M. Rissi[c], A. Robert[e], G. E. Romero[v], S. Rügamer[a], A. Saggion[h], A. Sánchez[e], P. Sartori[h], V. Scalzotto[h], V. Scapin[h], R. Schmitt[a], T. Schweizer[m], M. Shayduk[m], K. Shinozaki[g], S. N. Shore[s], N. Sidro[b†], A. Sillanpää[l], D. Sobczynska[j], A. Stamerra[o], L. S. Stark[c], L. Takalo[l], P. Temnikov[q], D. Tescaro[b], M. Teshima[g], N. Tonello[g], A. Torres[e], D. F. Torres[b,t], N. Turini[o], H. Vankov[q], V. Vitale[i], R. M. Wagner[g], T. Wibig[j], W. Wittek[g], R. Zanin[h], J. Zapatero[e]

[a] Universität Würzburg, D-97074 Würzburg, Germany
[b] Institut de Física d'Altes Energies, Edifici Cn., E-08193 Bellaterra (Barcelona), Spain
[c] ETH Zurich, CH-8093 Switzerland
[d] Universidad Complutense, E-28040 Madrid, Spain
[e] Universitat Autónoma de Barcelona, E-08193 Bellaterra, Spain
[f] Universität Dortmund, D-44227 Dortmund, Germany
[g] Max-Planck-Institut für Physik, D-80805 München, Germany
[h] Università di Padova and INFN, I-35131 Padova, Italy
[i] Università di Udine, and INFN Trieste, I-33100 Udine, Italy
[j] University of Lodz, PL-90236 Lodz, Poland
[k] Yerevan Physics Institute, AM-375036 Yerevan, Armenia
[l] Tuorla Observatory, FI-21500 Piikkiö, Finland
[m] Humboldt-Universität zu Berlin, D-12489 Berlin, Germany
[n] University of California, Davis, CA-95616-8677, USA
[o] Università di Siena, and INFN Pisa, I-53100 Siena, Italy
[p] Università di Trieste, and INFN Trieste, I-34100 Trieste, Italy
[q] Institute for Nuclear Research and Nuclear Energy, BG-1784 Sofia, Bulgaria
[r] Osservatorio Astronomico and INFN Trieste, I-34100 Trieste, Italy
[s] Università di Pisa, and INFN Pisa, I-56126 Pisa, Italy
[t] Institut de Ciències de l'Espai, E-08193 Bellaterra (Barcelona), Spain
[u] Universitat de Barcelona, E-08028 Barcelona, Spain
[v] Instituto Argentino de Radioastronomia CC5, CP 1894 Villa Elisa, Argentina
[w] Cavendish Laboratory, University of Cambridge, CB3 0HE, UK.
[x] deceased
[†] to whom correspondence should be addressed. E-mail: nsidro@ifae.es, jrico@ifae.es

**Microquasars are binary star systems with relativistic radio-emitting jets. They are potential sources of cosmic rays and laboratories for elucidating the physics of relativistic jets. Here we report the detection of variable gamma-ray emission above 100 gigaelectron volts from the microquasar LS I +61 303. Six orbital cycles were recorded. Several detections occur at a similar orbital phase, suggesting the emission is periodic. The strongest gamma-ray emission is not observed when the two stars are closest to one another, implying a strong orbital modulation of the emission or the absorption processes.**

Microquasars are binary star systems consisting of a compact object of a few solar masses (either a neutron star or a black hole), and a companion star that loses mass into an accretion disk around the compact object. The most relevant feature of microquasars is that they display relativistic jets, outflows of matter from regions close to accreting black holes and neutron stars, which remain among the most spectacular, yet poorly explained astrophysical phenomena (*1*). In microquasars, the time scales of relevant processes, being correlated with the compact object mass, are reduced millions of times in comparison with quasars, allowing the study of a different range of variability. In addition, microquasars could measurably contribute to the density of Galactic cosmic rays (*2*). It is worth noticing that photons up to Very High Energy (VHE) are an expected by-product of cosmic ray production.

Two GeV sources detected by the Energetic Gamma-Ray Experiment Telescope (EGRET) (*3*) are positionally compatible with microquasars. One of these, LS 5039 (*4*), in the southern hemisphere, has been recently confirmed as a TeV emitter (*5*) by the High Energy Stereoscopic System (H.E.S.S.). The second one, LS I +61 303 (*6*), is at the northern hemisphere, and thus a natural target for the Major Atmospheric Gamma-ray Imaging Cherenkov (MAGIC) telescope.

LS I +61 303 is a B0 main sequence star with a circumstellar disc, i.e. a Be star, located at a distance of ~2 kpc (*7*). A compact object of unknown mass is orbiting around it every 26.496 days (*8*, *9*). The eccentricity of the orbit is 0.72±0.15 and the periastron passage (the point where the two stars are closest to one another) is at phase 0.23±0.02. The nature of the compact object is still debated (*10*). Radio outbursts are observed every orbital cycle from this system (*11*) at phases varying between 0.45 and 0.95 (*12*) with a modulation of 4.6 years. The monitoring of LS I +61 303 at X-ray energies (*13*-*15*) revealed X-ray outbursts starting at around phase 0.4 and lasting up to phase 0.6. The detection of extended jet-like and likely rapidly precessing radio emitting structures at angular extensions of 0.01-0.05 arcsec has been interpreted as an unambiguous evidence of the microquasar nature of LS I +61 303 (*16*, *17*).

The gamma-ray source 2CG 135+01 (also 3EG J0241+6103) was discovered by the COsmic ray Satellite B (COS-B) at energies above 100 MeV (*18*). Despite its large uncertainty (~1°) in position, the source was early proposed to be the high-energy counterpart of the 26.5-day periodic radio outburst source GT 0236+610, which turned out to be the early type star LS I +61 303 (*11*). The large uncertainty of the position of 3EG J0241+6103/2CG 135+01 did not allow unambiguous association with LS I +61 303. The GeV gamma-ray emission from this EGRET source is clearly variable (*19*). Even though the GeV data are yet scarce in this regime, an increased emission has been hinted at for the periastron passage (*20*), and firmly reported around phase 0.5 (*6*), coincident with the X-ray outbursts.

LS I +61 303 was observed using MAGIC, located on the Canary Island of La Palma (Spain). MAGIC is an Imaging Air Cherenkov Telescope (IACT). This kind of instruments images the Cherenkov light produced in the particle cascade initiated by a gamma-ray in the atmosphere. MAGIC (*21*, *22*) includes several innovative techniques and technologies in its design, and is currently the largest single dish telescope (17 m diameter) in this energy band. It is equipped with a 576-pixel, 3.5° field-of-view photomultiplier camera. The telescope's sensitivity above 100 GeV is about 2.5% of the Crab Nebula flux (the calibration standard candle for IACTs) in 50 hours of observations. The angular resolution is about 0.1° and the energy resolution above 200 GeV better than 30%. MAGIC can provide gamma-ray source localization in the sky with a precision of ~2′.

LS I +61 303 was observed during 54 hours (after standard quality selection, discarding bad weather data) between October 2005 and March 2006. MAGIC is unique among IACTs in its capability to operate in the presence of the Moon. This allows the duty cycle to be increased by up to 50%, thus considerably improving the sampling of variable sources. In particular, 22% of the data used in this analysis were recorded under moonlight. The data analysis was carried out using the standard MAGIC analysis and reconstruction software (*21*, *22*).

The reconstructed gamma-ray map (Fig. 1) during two different observation periods, around periastron passage and at higher (0.4-0.7) orbital phases, clearly shows an excess in the latter case. The excess is located at (J2000): $\alpha = 2^h40^m34^s$, $\delta = 61°15'25''$, with statistical and systematic uncertainties of ±0.4′ and ±2′, respectively, in agreement with the position of LS I +61 303. The distribution of gamma-ray excess is consistent with a point-like source. In the natural case in which the VHE emission is produced by the same object detected at EGRET energies, this result identifies a gamma-ray source that resisted classification during the last three decades.

Our measurements show that the VHE gamma-ray emission from LS I +61 303 is variable. The gamma-ray flux above 400 GeV coming from the direction of LS I +61 303 (Fig. 2) has a maximum corresponding to about 16% of the Crab Nebula flux, and is detected at around phase 0.6. The combined statistical significance of the 3 highest flux measurements is $8.7\sigma$, for an integrated observation time of 4.2 hours. The probability for the distribution of measured fluxes to be a statistical fluctuation of a constant flux (obtained from a $\chi^2$ fit of a constant function to the entire data sample) is $3\times10^{-5}$. The fact that the detections occur at similar orbital phases hints at a periodic nature of the VHE gamma-ray emission. Contemporaneous radio observations of LS I +61 303 were carried out at 15 GHz with the Ryle Telescope covering several orbital periods of the source. The peak of the radio outbursts was at phase 0.7, just after the increase of the VHE gamma-rays flux (Fig. S1).

The VHE spectrum derived from data between ~200 GeV and ~4 TeV at orbital phases between 0.4 and 0.7 is fitted reasonably well ($\chi^2$/ndf = 6.6/5) by a power law function: $dN_\gamma/(dA/dt/dE) = (2.7 \pm 0.4 \pm 0.8) \times 10^{-12} (E/TeV)^{(-2.6 \pm 0.2 \pm 0.2)}$ $cm^{-2}$ $s^{-1}$ $TeV^{-1}$. Errors quoted are statistical and systematic, respectively (Fig. S2). This spectrum is consistent with that of EGRET for a spectral break between 10 and 100 GeV. We estimate that the flux from LS I +61 303 above 200 GeV corresponds to an isotropic luminosity of ~7 x $10^{33}$ erg $s^{-1}$, at a distance of 2 kpc. The intrinsic luminosity of LS I +61 303 at its maximum is a factor ~6 higher than that of LS 5039, and a factor ~2 lower than the combined upper limit (<8.8 x $10^{-12}$ $cm^{-2}$ $s^{-1}$ above 500 GeV) obtained by Whipple (*23*). LS I +61 303 displays more luminosity at GeV than at X-ray energies, a behavior shared also by LS 5039 (*4*).

Different models have been put forward to explain putative gamma-ray emission from LS I +61 303. Maraschi & Treves (*24*) early proposed that the GeV emission measured by COS-B could be due to the wind from a pulsar interacting with that of the stellar companion. However, the detection of radio jets (*16*, *17*) favored an accretion scenario. The observation of jets has also triggered the study of different microquasar gamma-ray emission models, some based on hadronic mechanisms: relativistic protons in the jet interact with non-relativistic stellar wind ions, producing gamma-rays via neutral pion decay (*25*, *26*); some based on leptonic mechanisms: inverse Compton (IC) scattering of relativistic electrons in the jet on stellar and/or synchrotron photons (*27-29*).

The TeV flux maximum is detected at phases 0.5-0.6 (Fig. 2), overlapping with the X-ray outburst and the onset of the radio outburst. The maximum flux is not detected at periastron, when the accretion rate is expected to be the largest (*30*). This result seems to favor the leptonic over the hadronic models, since the IC efficiency is likely to be higher than that of proton-proton collisions at the relatively large distances from the companion star at this orbital phase. Concerning energetics, a relativistic power of several $10^{35}$ erg $s^{-1}$ could explain the non thermal luminosity of the source from radio to VHE gamma-rays. This power can be extracted from accretion in a slow inhomogeneous wind along the orbit (*30*).

The variable nature of the TeV emission on timescales of ~1 day constrains the emitting region to be smaller than a few $10^{15}$ cm (or ~0.1 arcsec at 2 kpc). This is compatible with the emission being produced within the binary system, where there are large densities of seed photons for IC interaction. Under these strong photon fields opacity effects certainly play a role in the modulation of the emitted radiation (*31*). Indeed VHE gamma-ray emission that peaks after periastron passage has been recently predicted considering electromagnetic cascading within the binary system (*32*).

In addition, the detection of VHE gamma-ray emission associated with both LS I +61 303 and LS 5039, observationally confirms high-mass microquasars as a new population of Galactic TeV gamma-ray sources.

LS I +61 303 is an excellent laboratory to study the VHE gamma-ray emission and absorption processes taking place in massive X-ray binaries: the high eccentricity of the binary system provides very different physical conditions to be tested on timescales of less than one month. Future MAGIC observations will test both, the periodicity of the signal and its intra-night variability.

34. We would like to thank the IAC for the excellent working conditions at the Observatory de los Muchachos in La Palma. The support of the German BMBF and MPG, the Italian INFN and the Spanish CICYT is gratefully acknowledged. This work was also supported by ETH Research Grant TH-34/04-3 and the Polish MNiI Grant 1P03D01028.

**Supporting Online Material**

www.sciencemag.org

Figs. S1, S2

**Figures**

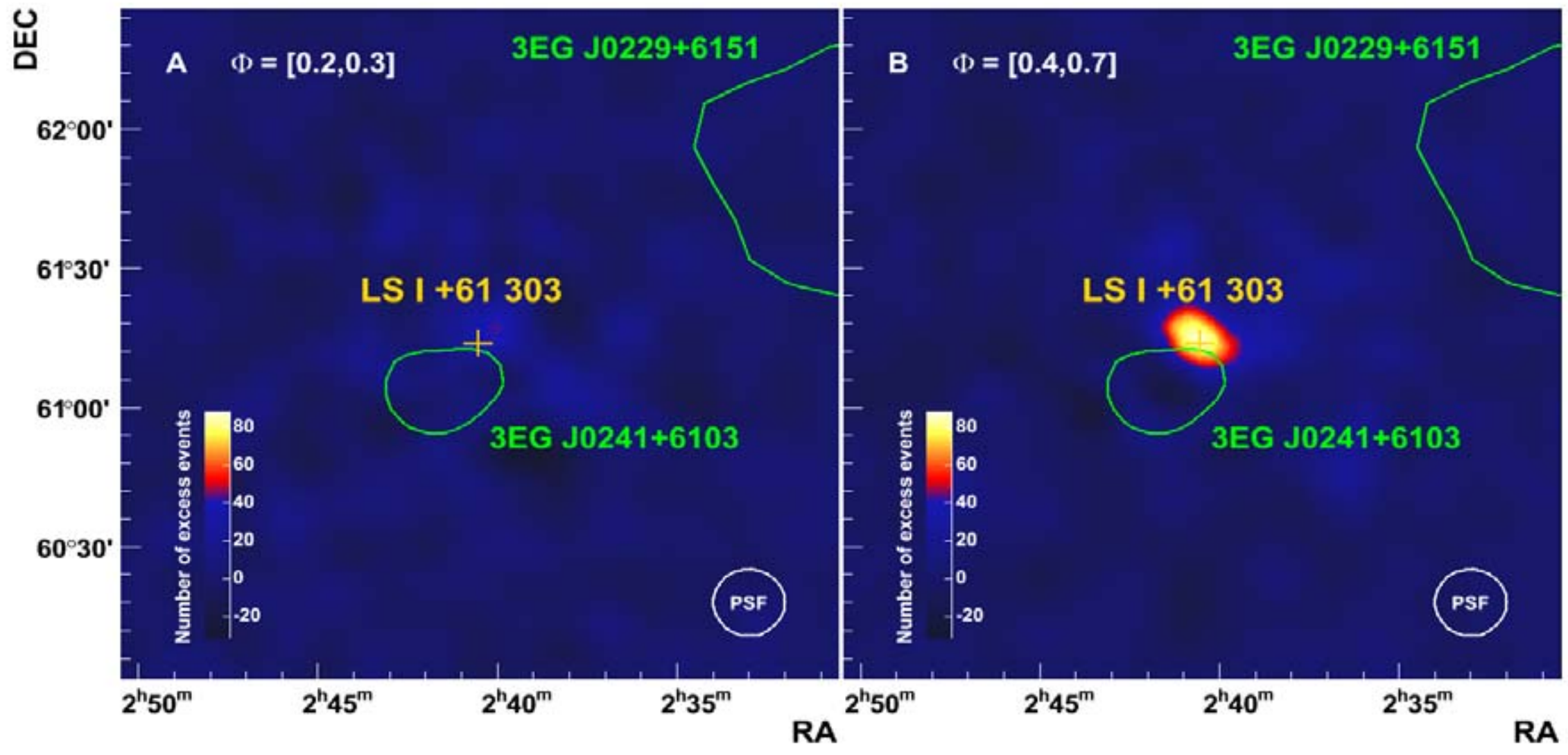

**Fig. 1.** Smoothed maps of gamma-ray excess events above 400 GeV around LS I +61 303. (A) 15.5 hours corresponding to data around periastron, i.e. between orbital phases 0.2 and 0.3. (B) 10.7 hours at orbital phase between 0.4 and 0.7. The number of events is normalized in both cases to 10.7 hours of observation. The position of the optical source LSI +61 303 (yellow cross) and the 95% confidence level (CL) contours for 3EG J0229+6151 and 3EG J0241+6103 (green contours), are also shown. The bottom-right circle shows the size of the point spread function of MAGIC (1σ radius). No significant excess in the number of gamma-ray events is detected around periastron passage, while it shows up clearly (9.4σ statistical significance) at later orbital phases, in the location of LS I +61 303.

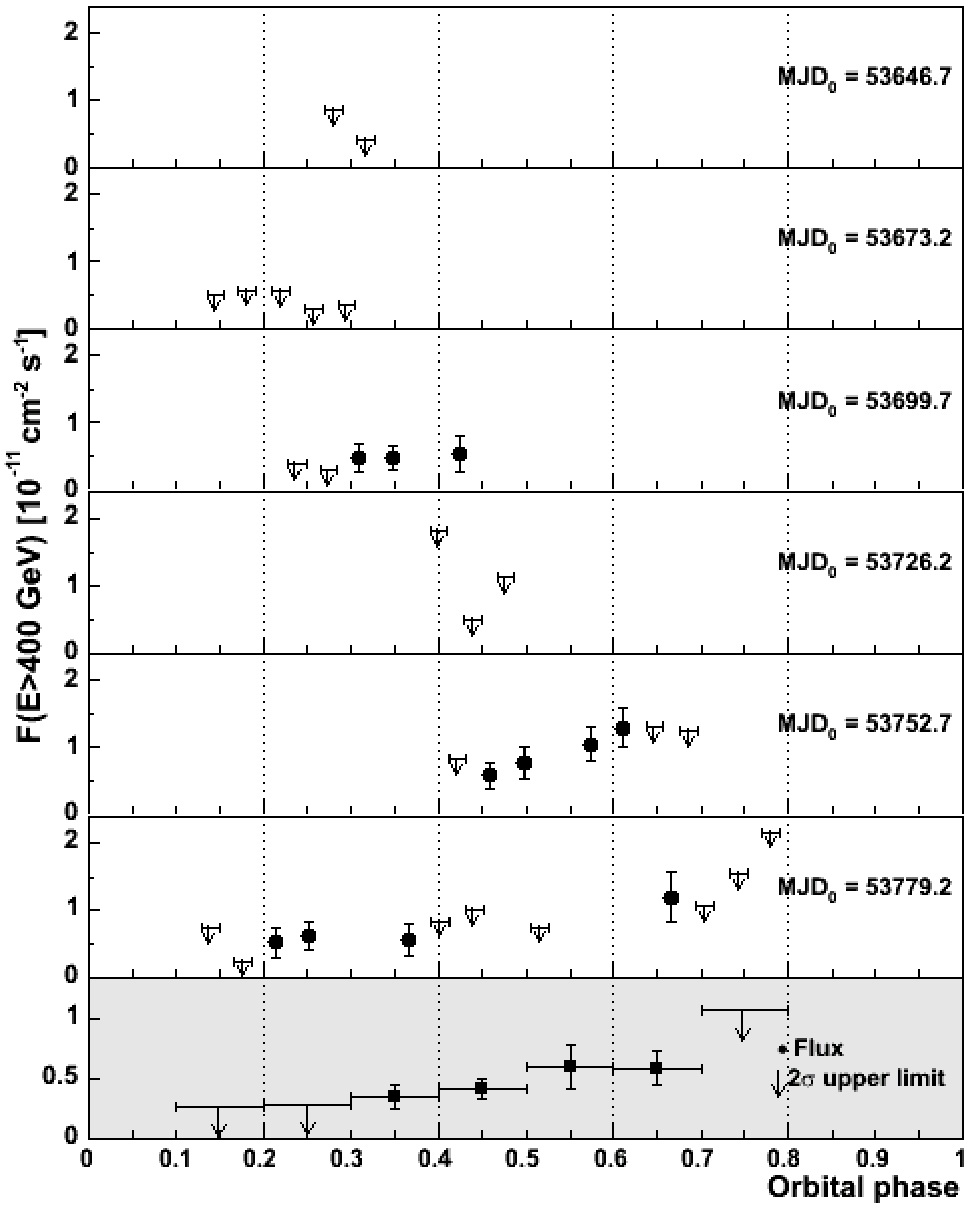

**Fig. 2.** VHE gamma-ray flux of LS I +61 303 as a function of the orbital phase for the six observed orbital cycles (6 upper panels, one point per observation night), and averaged for the entire observation time (lowermost panel). Vertical error bars include 1σ statistical error and 10% systematic uncertainty on day-to-day relative fluxes. Only data points with more than 2σ significance are shown, and 2σ upper limits (*33*) are derived for the rest. The Modified Julian Date (MJD) corresponding to orbital phase 0 is indicated for every orbital

cycle. The orbital phase is computed with $P_{orb}$=26.4960 d and $T_0$=JD 2443366.775 (*9*); periastron takes place at phase 0.23 (*10*). Marginal detections occur between orbital phases 0.2 and 0.4 in different cycles, while a significant increase of flux is detected from phase ~0.45 to phase ~0.65 in the 5th cycle, peaking at ~16% of the Crab Nebula flux on MJD 53769 (phase 0.61). During the following cycle, the highest flux is measured on MJD 53797 (phase 0.67). This behavior suggests that the VHE gamma-ray emission from LS I +61 303 has a periodic nature.

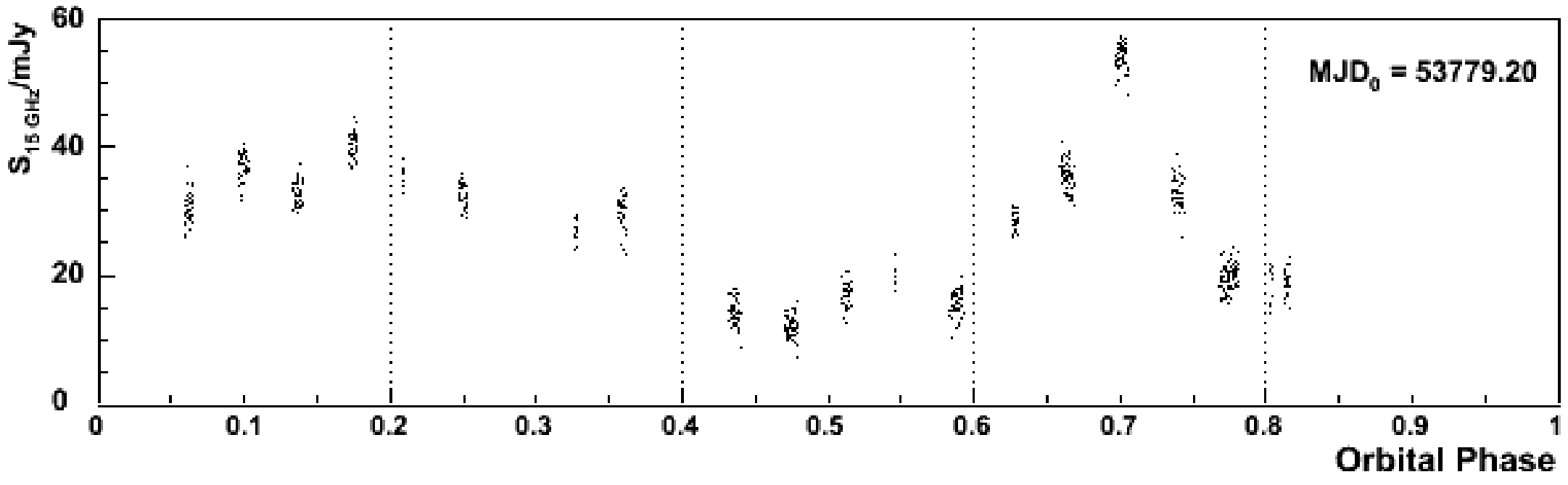

**Fig. S1.** LS I +61 303 radio flux density at 15 GHz measured with the Ryle Telescope for the last orbital cycle observed by MAGIC (from 14 February to 8 March 2006). The day corresponding to orbital phase 0 is indicated. The periodic radio outburst has its maximum at MJD 53798.8, corresponding to an orbital phase of 0.70.

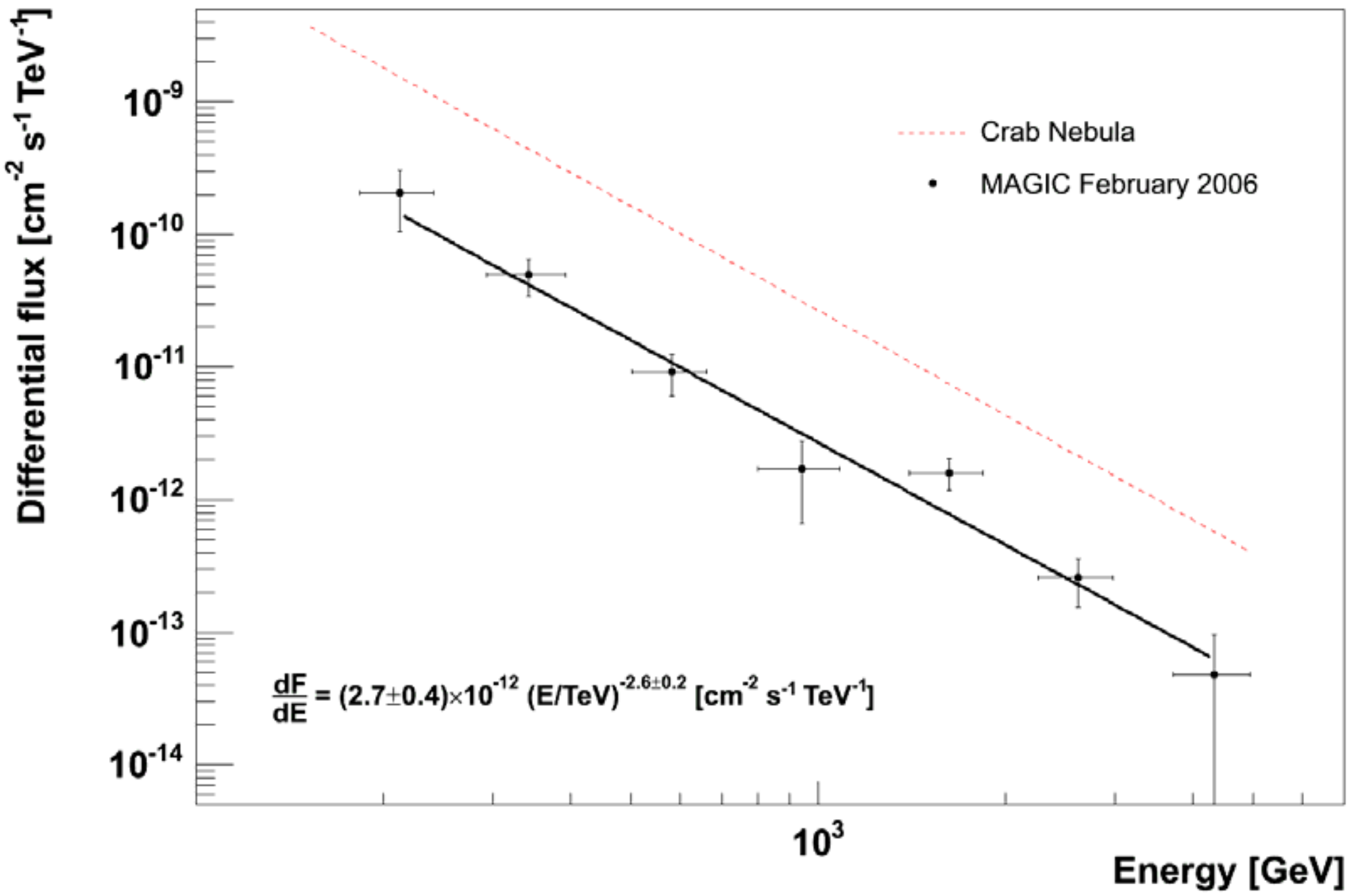

**Fig. S2.** Differential energy spectrum measured for LS I +61 303 for orbital phases between 0.4 and 0.7 and energies between 200 GeV and 4 TeV. The error bars show the 1$\sigma$ statistical uncertainty. The dashed, red line corresponds to the Crab Nebula differential spectrum measured by MAGIC. The solid, black line is a fit of a power law (also expressed mathematically in the inset) to the measured points.